\documentclass{article}

\usepackage{microtype}
\usepackage{graphicx}
\usepackage{subfigure}
\usepackage{booktabs} 

\usepackage{hyperref}

 \usepackage[accepted]{icml2023}

\usepackage{amsmath}
\usepackage{amssymb}
\usepackage{mathtools}
\usepackage{amsthm}

\usepackage[capitalize,noabbrev]{cleveref}

\theoremstyle{plain}

\theoremstyle{definition}

\theoremstyle{remark}

\usepackage[textsize=tiny]{todonotes}

\icmltitlerunning{Neural Astrophysical Wind Models}

\begin{document}

\twocolumn[
\icmltitle{Neural Astrophysical Wind Models}



\icmlsetsymbol{equal}{*}

\begin{icmlauthorlist}
\icmlauthor{Dustin Nguyen}{yyy}
\end{icmlauthorlist}

\icmlaffiliation{yyy}{Department of Physics, The Ohio State University, Columbus, Ohio, USA}

\icmlcorrespondingauthor{Dustin Nguyen}{dnguyen.phys@gmail.com}

\icmlkeywords{Machine Learning, ICML}

\vskip 0.3in
]



\printAffiliationsAndNotice{}  

\begin{abstract}
The bulk kinematics and thermodynamics of hot supernovae-driven galactic winds is critically dependent on both the amount of swept up cool clouds and non-spherical collimated flow geometry. However, accurately parameterizing these physics is difficult because their functional forms are often unknown, and because the coupled non-linear flow equations contain singularities. We show that deep neural networks embedded as individual terms in the governing coupled ordinary differential equations (ODEs) can robustly discover both of these physics, without any prior knowledge of the true function structure, as a supervised learning task. We optimize a loss function based on the Mach number,  rather than the explicitly solved-for 3 conserved variables, and apply a penalty term towards near-diverging solutions. The same neural network architecture is used for learning both the hidden mass-loading and surface area expansion rates. This work further highlights the feasibility of neural ODEs as a promising discovery tool with mechanistic interpretability for non-linear inverse problems.

\end{abstract}

\section{Introduction}
Mathematics and the physical laws that have followed suit are unreasonably effective in creating models with predictive power for the natural sciences \citep{Wigner1960}. In physics, understanding implies that the governing principles are written in the form of explicit differential equations of motion \citep{Meiss2007}. However, when modelling real world systems, we do not always know what the true underlying physics is.

In the context of explaining observations of nearby galactic superwinds, recent works have suggested the inclusion of additional physics can help overcome the so-called ``cloud-crushing problem" \citep{Klein1994}, which challenges the existence of observed high velocity cool clouds. These efforts include incorporation of efficient multi-phase mixing \citep{Gronke2018,Gronke2020}, cosmic ray acceleration \citep{Quataert2022a,Quataert2022b}, among many other mechanisms \citep{Faucher-Giguere2023}. While incorporation of new physics into semi-analytic models and 3D simulations are a vibrant and active field of research \citep[e.g.,][and many others]{Schneider2015,Kim2020,Pandya2022,Wibking2022,Smith2023,Nguyen2023,Tan2023}, the development of optimization methods to systematically discover and characterize the structure of physics from observational data has not received similar attention. Namely, descriptions of superwinds from nearby starburst prototype M82 have relied on parameter estimation of classic theoretical models \citep{Chevalier1985,Strickland2009,Lopez2020,Nguyen2021,Yuan2023}. Methods for developing new data-driven models have not yet been explored for such systems. 

Deep neural networks \citep{LeCun2015,Goodfellow2016} can be used as a universal approximator for unknown functions and operators \citep{HORNIK1990,Lu2021}. Recently, deep neural networks have been used as numerical discretization schemes of ODEs/PDEs \citep{Chen2019} based on the adjoint sensitivity method. This idea has been expanded to a hybrid model termed the Universal Differential Equation \citep[UDE,][]{rackauckas2020universal} that allows one to freely augment ODEs/PDEs with universal approximators such as neural networks. Embedded neural networks are guaranteed to obey conservation laws, as they can only influence the dynamics of the system term by term. In this paper, neural ODEs are ODEs with a neural network as a part of the equations and are solved using standard numerical methods (unlike Physics-Informed Neural Networks, where the numerical solver is a ``black box"). Neural networks embedded within ODEs/PDEs are a new and active field of research and have been used to predict COVID-19 pandemic waves \citep{Kuwahara2023}, and various other applications within biology, chemistry, mathematics, and physics \citep{Vortmeyer-kley2021,Gelbrecht2021,Keith2021, Fronk2023,Stepanmiants2023,Yin2023,Santana2023}. 

\newpage

We present the first analysis of neural coupled ODEs, within the UDE framework, that contain singularities. To illustrate generality, we consider two experiments where the embedded neural network learns two different types of physics: 1. mass-loading of swept up clouds, and, 2. the surface area expansion rate attributed to flow geometry. We define a custom loss function based solely on the Mach number, rather than the 3 individual conserved variables, and adopt linear scaling towards early integration steps. Additionally, this loss function penalizes solutions that approach the sonic point, which is infamously known to lead to numerical instability \citep{Lamers1999}. We will show that regression on such a loss function is sufficient to characterize the underlying $v$, $\rho$, and $P$ profiles and discover the hidden physics embedded within the training data. 

\section{Methods}
\begin{figure}
    \centering
    \includegraphics[width=0.6\columnwidth]{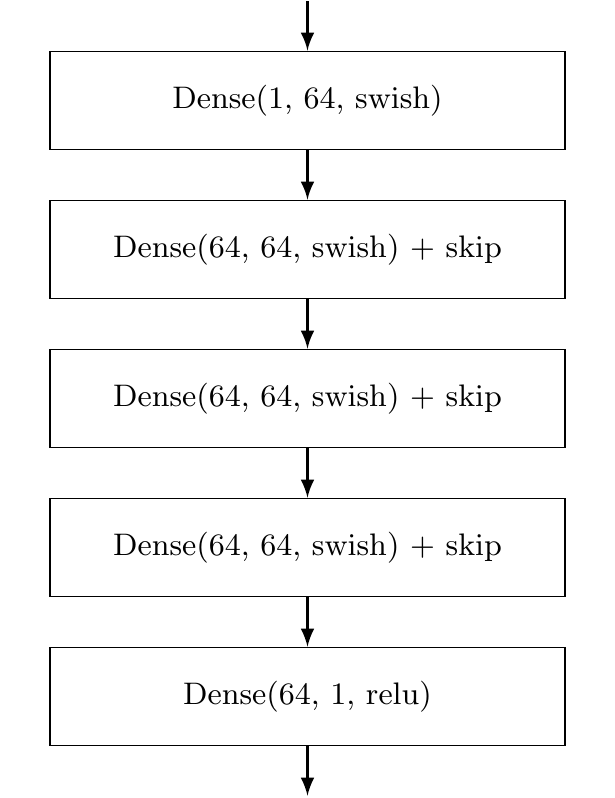}
    \caption{Architecture of neural network $\Phi(x)$. The input is a single position $x$, which then gets passed through a total of 5 \texttt{Dense} layers where the three hidden layers contain skip connections and 64 nodes. Each layer uses \texttt{swish} as the activation function, with the output layer using \texttt{relu} to ensure positivity.}
    \label{fig:neural_network}
\end{figure}

The hydrodynamic equations for a non-radiative highly supersonic steady-state hot flow moving in the $x$ direction are \citep{Cowie1981,Nguyen2021}
\begin{align}
    & \frac{1}{A} \frac{d}{dx} ( \rho v A ) = \dot{\mu} , \label{eq:continuity}   \\ 
    & v \frac{dv}{dx} = - \frac{1}{\rho} \frac{dP}{dx}  - \frac{\dot{\mu}v}{\rho} ,  \label{eq:momentum} \\
    & v \frac{d\epsilon}{dx} - \frac{v P}{\rho ^2 } \frac{d\rho}{dx} =  \frac{\dot{\mu}}{\rho} \bigg( \frac{v^2}{2} - \epsilon - \frac{P}{\rho} \bigg), \label{eq:energy} 
\end{align}
where $v$, $\rho$, $P$, $\epsilon$, $\dot{\mu}$, and $A(x)$ are the bulk velocity, density, pressure, specific internal energy, volumetric mass-loading rate, and flow area, respectively. By substituting the equations into each other, the derivatives will each contain the singularity: $(\mathcal{M}^2 - 1)^{-1}$ \citep[i.e., is a pole at the sonic point where $\mathcal{M}=1$. See ][]{Lamers1999}. We consider two experiments of training a neural network to learn: 1. the volumetric mass-loading rate, $\dot{\mu}$, and 2. the surface area expansion rate $ d \, \ln A / dx$. In each, the training dataset is calculated with the analytic functional form of either $\dot{\mu}(x)$ or $ d \, \ln A / dx$. After generating the training dataset, we forget the true physical variable and replace it with neural network  $\Phi(x)$ that has a total of 5 layers (See Fig. \ref{fig:neural_network} for details). 

The dimensions of each simulation is $0.3\,\mathrm{kpc} \leq x \leq 2\mathrm{kpc}$ and number of steps $n_x = 500$. We have assumed the flow to have already left the starburst volume $R=0.3\,$kpc  \citep{Chevalier1985}. The initial conditions are then $v_0 = 1500 \, \mathrm{km\,s^{-1}}$, $n_0 = 10 \, \mathrm{cm^{-3}}$, and $T_0 = 10^7\,\mathrm{K}$, which corresponds to a highly supersonic $(\mathcal{M}_0 \sim 3)$ non-radiative wind that has asymptotic velocities roughly within the allowed range for starburst galaxy M82 \citep{Strickland2009}. 

We calculate the loss function using the Mach number $\mathcal{M} = v / (\gamma P/\rho)^{1/2}$. We find this method to outperform summing Mean-Squared-Error (MSE) of the individual conserved variables $v$, $\rho$, and $P$, even when using min-max normalization. The loss function is 
\begin{equation}
    \mathcal{L} = \sum_i^{n_x} \bigg[ \mathcal{W}_i \times \bigg(\mathcal{M}_i - \hat{\mathcal{M}}_i\bigg)^2 \bigg] + \zeta_\mathrm{penalty} (\mathcal{M}) 
    \label{eq:loss_function}
\end{equation}
The first term is a weighted MSE, where the weights $\mathcal{W}_i$ linearly scale the MSE as a function of $n_x$, where $\mathcal{W}_0=1$ and $\mathcal{W}_{n_x}=\kappa < 1$. We do not include division by $n_x$, as it does not impact training. We vary $\kappa$ during training for each optimization algorithm (details below). This scaling increases sensitivity to early solutions, which is important for non-linear problems. The latter term $\zeta_\mathrm{penalty}(\mathcal{M})$ penalizes solutions with Mach numbers close to $1$, as to prevent the neural network from sampling diverging solutions ($\mathcal{M} = 1 $) and is a function of the total loss $\mathcal{L}_j$ and $\mathcal{M}_i$ as: 
\begin{align}
    & \zeta_{\mathrm{penalty},j} = \mathcal{L}_j \times \omega \sum_i^{nx}\bigg[1-\bigg(1-\hat{\mathcal{M}}_i\bigg)^2 \bigg]   \\ & \quad \quad \quad \quad \quad \quad  (\mathcal{M}_\mathrm{min} \leq \hat{\mathcal{M}}_i \leq \mathcal{M}_\mathrm{max} ) , \label{eq:penalty_term} 
\end{align}
where $j$ is the training iteration, $\omega$ is the penalty weight, $\mathcal{M}_\mathrm{min} = 1$, and $\mathcal{M}_\mathrm{max}=1.5$ sets the threshold for penalization. We summarize the training process in the pseudo-code below. 

\begin{algorithm}[tb]
   \caption{Training Psuedo-Code}
   \label{alg:example}
\begin{algorithmic}
   \STATE {\bfseries Input:} Initial Condition Array $(v_0,\rho_0,P_0)$ size $1 \times 3$
   \STATE \textbf{1. } Calculate true solution using $\dot{\mu}(x)$ or $A(x)$ in Eqs.~1,2,3. The solution is then a matrix $y$ of size $n_x \times 3$. Use $y$ to calculate $\mathcal{M}$ which is size $n_x \times 1$. 
   \STATE \textbf{2. } Calculate prediction using neural network $\Phi(x)_{p_0}$ within neural equations. $p_0$ is initial parameters array for constructing $\Phi(x)_{p_0}$. The solution is a matrix $\hat{y}_0$ of size $n_x \times 3$. Use $\hat{y}_0$ to calculate $\hat{\mathcal{M}_0}$ which is size $n_x \times 1$. 
   \FOR{$j=0$ {\bfseries to} $n_\mathrm{iters}$}
   \STATE Reconstruct neural net $\Phi(x)_{p_j}$ using parameters $p_j$.
   \STATE Calculate $\hat{y}_j$, then $\hat{\mathcal{M}}_j$.
   \STATE Calculate loss $\mathcal{L}_j$ using MSE between $\mathcal{M}$ and $\hat{\mathcal{M}}_j$ (batched over all positions $n_x$)
   \IF{$\mathcal{M}_\mathrm{min} \leq \hat{\mathcal{M}}_j \leq \mathcal{M}_\mathrm{max}$ (batch)}
   \STATE $\mathcal{L}_j += \mathcal{L}_{j} * \mathrm{Penalty \ Term} $ 
   \ENDIF
   \STATE Optimize for new set of parameters $p_j$ that minimize $\mathcal{L}_j$ using \texttt{ADAM} or $\texttt{BFGS}$ optimization algorithms. 
   \ENDFOR{ \textbf{if} $j=n_\mathrm{iters}$ \textbf{or} gradient tolerance met}
\end{algorithmic}
\end{algorithm}

We solve the equations (both classical and neural) using standard numerical methods involving 5th order RK4 \citep{TSITOURAS2011} with adaptive step size control \citep{rackauckas2017differentialequations}. We calculate gradients using forward-mode automatic differentiation \citep{RevelsLubinPapamarkou2016}. All of the work is done using packages contained within \texttt{Julia}'s Scientific Machine Learning (SciML) ecosystem \citep{rackauckas2020universal}.  

\section{Results}

\textbf{Learning Mass-loading:} We test the neural network's ability to learn mass-loading $\dot{\mu}(x) = \dot{\mu} (\lambda / x )^\Delta$. This represents intense cloud entrainment immediately after the wind leaves the host galaxy ($\lambda = 0.3\,\mathrm{kpc}$, $\Delta=4$, and $\dot{\mu}_0 = 500\,\mathrm{M_\odot \, yr^{-1} \, kpc^{-3}}$). Here we take the flow geometry to be spherical (i.e., $A(x) \propto x^2$). After calculating the training (true) solution with this function, we ``forget" $\dot{\mu}(x)$ and replace it with a neural network $\Phi(x)$ (See Fig.~\ref{fig:neural_network}) such that Equations \ref{eq:continuity},\,\ref{eq:momentum},\,and\,\ref{eq:energy} are re-written as neural wind equations:  
\begin{align}
    & \frac{1}{x^2} \frac{d}{dx} ( \rho v x^2 ) = \Phi , \label{eq:neural_mudot_continuity}   \\ 
    & v \frac{dv}{dx} = - \frac{1}{\rho} \frac{dP}{dx}  - \frac{\Phi v}{\rho} ,  \label{eq:neural_mudot_momentum} \\
    & v \frac{d\epsilon}{dx} - \frac{v P}{\rho ^2 } \frac{d\rho}{dx} =  \frac{\Phi}{\rho} \bigg( \frac{v^2}{2} - \epsilon - \frac{P}{\rho} \bigg). \label{eq:neural_mudot_energy}
\end{align}
We initialize the neural network with an output of approximately 0 (using \texttt{Flux.jl}'s default \texttt{Glorot} initialization). This implies that the first prediction solution is identical to an adiabatic spherical wind with no mass-loading (i.e., $\dot{\mu} \simeq 0 $). We train for $n_\mathrm{iters} = 300$ iterations using \texttt{ADAM} with a learning rate of $0.01$ and exponential decay rate of $0.98$, followed by additional iterations using \texttt{BFGS}. We use $\kappa = 0.1$ for scaling solutions. We use penalty term $\omega = 0.1$ and $\omega=0.75$ for \texttt{ADAM} and \texttt{BFGS}, respectively. Optimization with \texttt{BFGS} converges to the set gradient tolerance of $10^{-9}$ after 304 iterations. 

\begin{figure*}
    \centering
    \includegraphics[width=\textwidth]{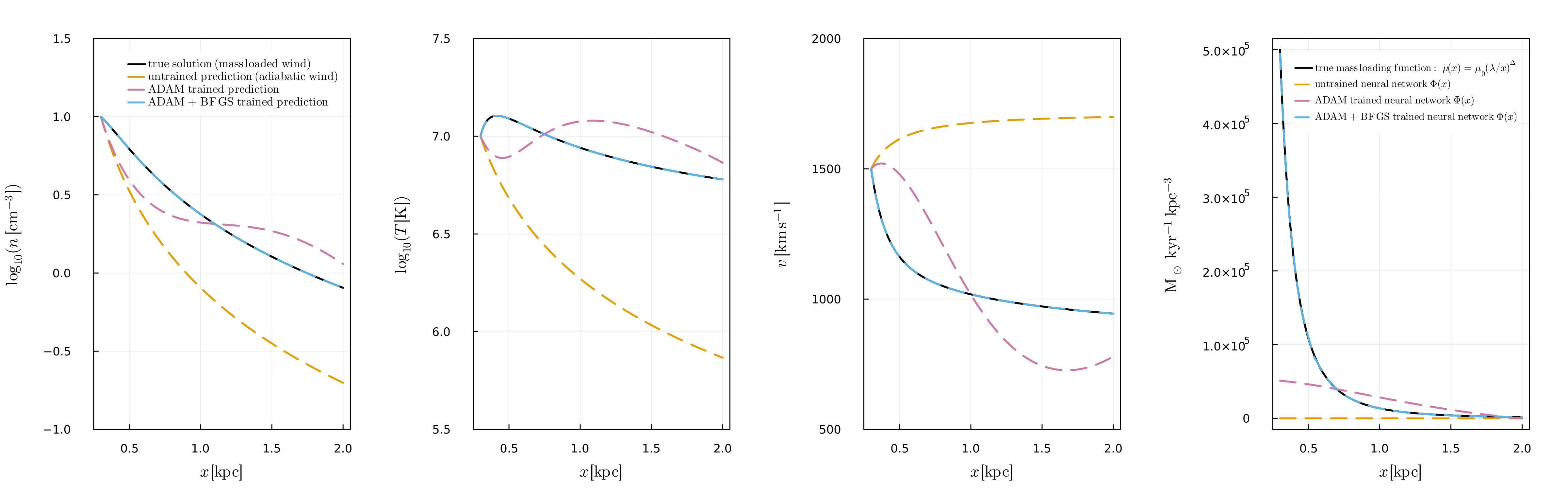}
    \caption{Density, temperature, and velocity profiles for solutions (left 3 panels) calculated with the mass-loading (right panel) given by the true function (black solid) and the output of neural networks (colored dashed). After \texttt{BFGS} and \texttt{ADAM} training, the three kinematic and thermodynamic profiles are well-matched (blue dashed on black lines), and the underlying mass-loading function was learned.}
    \label{fig:nTv_powerlaw}
\end{figure*}

\begin{figure}
    \centering
    \includegraphics[width=0.9\columnwidth]{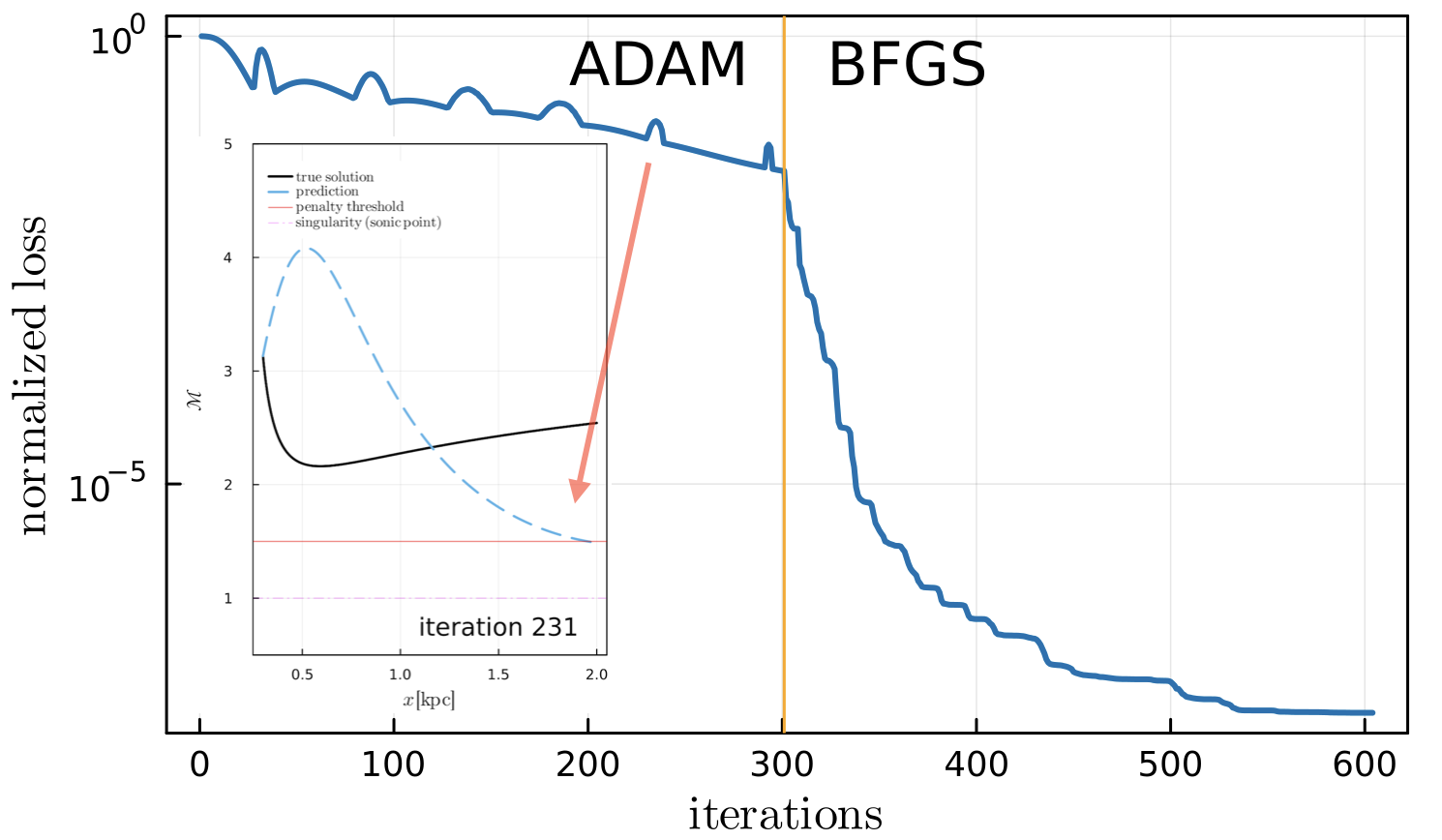}
    \caption{Learning curve for training neural network $\Phi(x)$ to learn underlying volumetric mass-loading rate $\dot{\mu}$ from the solved wind solutions. The bumps in the loss function during \texttt{ADAM} training epochs are attributed to penalties for near-diverging solutions.}
    \label{fig:loss_mudot}
\end{figure}

In the left three panels of Figure \ref{fig:nTv_powerlaw}, we plot three different solutions for velocity, number density ($n = \rho / \mu m_p)$, and temperature ($T = P / n k_B)$ radial profile predictions from: the true solution (solid black), the untrained model (orange dashed), the \texttt{ADAM} trained model (purple dashed), and the \texttt{ADAM} + \texttt{BFGS} trained (blue dashed). Although we explicitly train on only the Mach number, the three conserved variables ($v$, $\rho$, and $P$) end up getting well-matched (blue dashed lines). In the right most panel of Figure \ref{fig:nTv_powerlaw}, we plot various volumetric mass-loading rates calculated with the true function $\dot{\mu}=\dot{\mu}_0(\lambda/x)^\Delta$ (black solid) and the output of the neural networks (colored dashed). In Figure \ref{fig:loss_mudot}, we plot the learning curve. The bumps in the \texttt{ADAM} portion of the learning curve (first 300 iterations), demonstrate the optimizer sampling solutions within the penalty threshold, and then updating the mass-loading rate, encoded by $\Phi(x)$, for new predictions that move away from such penalties. Training decreases the normalized loss by nearly 7 orders of magnitude. The predictions of the neural ODEs agree with the true solutions, and the output of the neural network indicate that it has learned the true mass-loading function. 

\textbf{Learning Surface Area Expansion:} We now test the neural network's ability to learn properties of the flow geometry $A(x) = A_0 ( 1 + (x/\eta)^2)$, which describes a flared-cylinder flow tube that has previously been used for solar coronal hole flux tubes \citep{Kopp1976} and Milky-Way disk winds \citep{Everett2008}. Qualitatively, this function implies a roughly constant flow area up until $\eta$ when the flow begins to undergo spherical expansion. For illustration, Equation \ref{eq:continuity} can be expanded into differential form as: $d \ln \rho/dx + d \ln v/dx + d \ln A/dx = \dot{\mu}/ \rho v $. This means that the effects of flow geometry are governed specifically by the surface area expansion rate $d \, \ln A /dx$, and not the magnitude of $A(x)$ (e.g., flows expanding into $4\pi x^2$ or $2 \pi x^2$ behave the same). Setting $\dot{\mu}=0$ we generate the training dataset using $A(x) = A_0 ( 1 + (x/ \eta)^2)$ so that $d \ln A / dx = 2 x / (\eta^2  + x^2) $, where $A_0= 0.25\,\mathrm{kpc}$ and $\eta = 1.0\,$kpc. We now ``forget" $d \ln A / dx $, replacing it with neural network $\Phi(x)$. Again, we initialize the neural network with output of approximately 0. This implies that the first prediction solution is identical to an adiabatic planar wind (i.e., $A(x) \simeq \mathrm{constant}$). We train for 300 iterations using \texttt{ADAM} with a learning rate of $0.01$ and exponential decay rate of $0.98$, followed by additional iterations using \texttt{BFGS}. We take $\kappa = 0.1$ and $\omega = 0.1$. Optimization with \texttt{BFGS} converges to the set gradient tolerance after 47 iterations.

\begin{figure}
    \centering
    \includegraphics[width=\columnwidth]{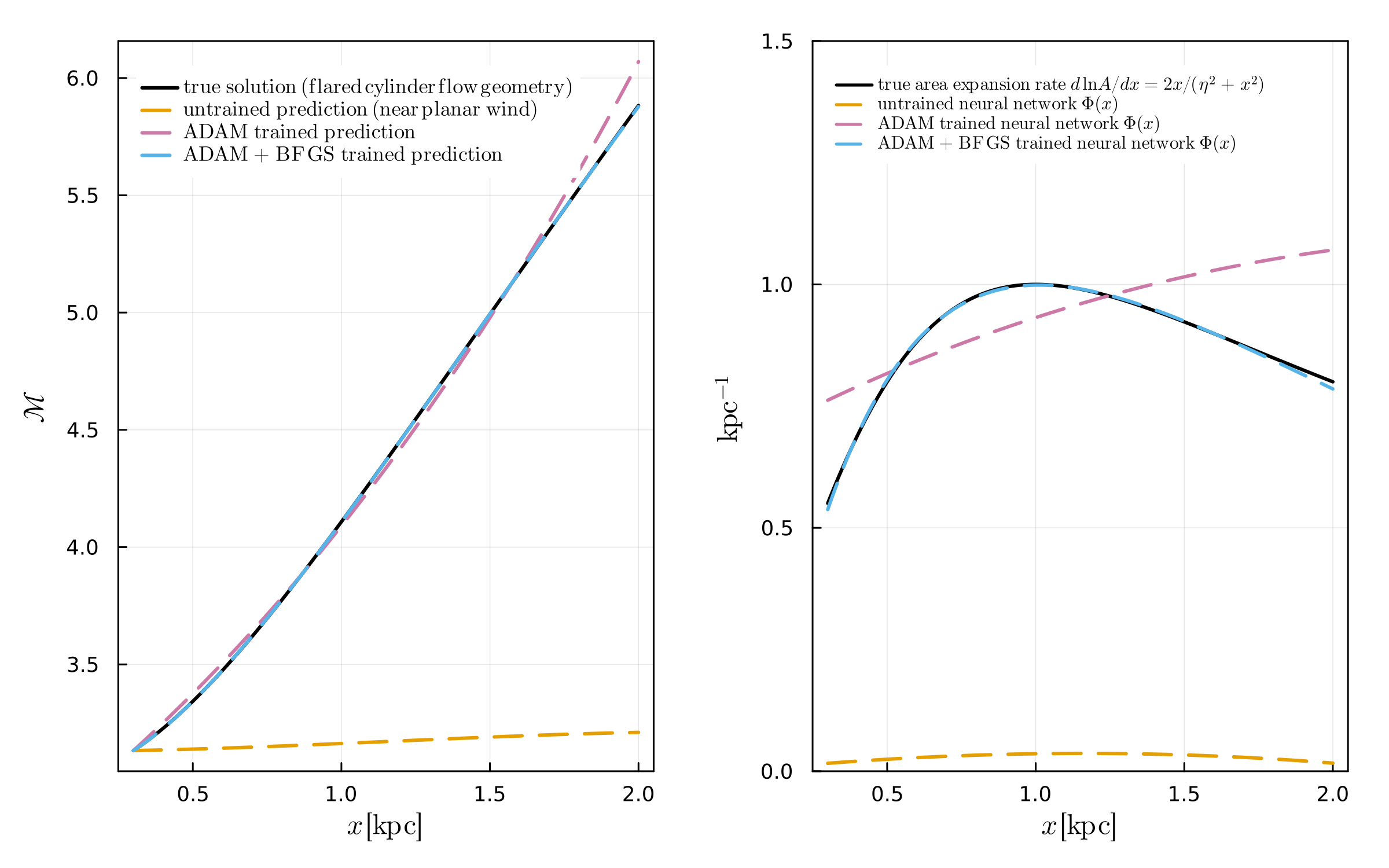}
    \caption{\textit{Left:} Mach number solutions calculated with the true area function (black solid), predictions from the neural ODEs (colored dashed lines). \textit{Right:} The true surface area expansion rate for a flared cylinder $d \log A  / dx$ (black solid) and the output from neural network $\Phi(x)$ for various degrees of training. }
    \label{fig:M_A_trained}
\end{figure}

In the left panel of Figure \ref{fig:M_A_trained}, we plot the Mach number solution calculated with the true area function (black line) and from predictions of the neural ODEs with varied degrees of training (colored dashed lines). We see that the best fitting line is, again, the model that underwent both $\texttt{ADAM}$ and $\texttt{BFGS}$ optimization. In the right panel we plot the surface area expansion rate calculated for the true surface area function (black), and the direct output of neural networks (colored dashed), finding agreement between the two (blue dashed and solid black).

\section{Conclusion and Future Work}
In this work we show that deep neural networks embedded as individual terms within astrophysical wind equations can discover the true functional form of both mass-loading and flow geometry expansion, without any prior knowledge. Additionally, to our knowledge at the time of writing, this is the first use of the UDE framework for coupled ODEs that contain singularities (i.e., poles at $\mathcal{M}=1$). To avoid diverging solutions, we introduced a penalty term that activates when the predicted solution falls within a threshold. This method is shown to be effective, and guides the optimization away from diverging solutions (Fig.~\ref{fig:loss_mudot}). We require \texttt{BFGS} optimization after \texttt{ADAM} to converge to the functional form of the true solution. $\texttt{BFGS}$ is a quasi-Newton method that uses second-order derivative information that outperforms $\texttt{ADAM}$ here at low values of normalized loss. We note that using \texttt{ADAM} at the beginning of training is still beneficial, as it is faster and can overcome local minima better. We demonstrate the same neural network architecture can be used to learn two very different types of physics. We show that the outputs of the neural network are interpretable (Figs.~\ref{fig:nTv_powerlaw}\,and\,\ref{fig:M_A_trained}). We find that training on the Mach number $\mathcal{M}$ outperformed training on the conserved variables $v$, $\rho$, and $P$. Beyond winds, we expect neural ODEs to be useful in discovering physics in other non-linear systems. 

In future work we would like to explore the predictive power of the trained neural networks, beyond the training dataset. Here, we specifically studied the ability of a neural network embedded within coupled ODEs to accurately capture physics during supervised training, and showcase the interpretability of the result. One possible direction is to use the SINDy \citep{Brunton2016} algorithm to extract symbolic representations of the trained neural networks which has been shown to be useful in creating hybrid models capable of making accurate long-term predictions for partially observed non-linear systems \citep{rackauckas2020universal,Santana2023,Fronk2023}. However, this approach requires the creation of a basis function library, which may not be generally applicable, especially when complicated non-polynomial functions are involved. If we continue to use the direct output of the trained neural network, careful considerations will need to be made towards: 1. more advanced neural network architectures that better  capture long-term dependencies such as LSTMs (as opposed to the simple forward-feed neural network used here), and 2. overfitting. Additionally, future work will investigate how noisy data may affect learning outcomes. We note that even in cases where predictive power beyond the training dataset is limited, the use of neural ODEs as a discovery tool for a contained dataset is still useful, as we demonstrate here.

\section*{Acknowledgements}

DN thanks Todd Thompson for useful discussions. DN also thanks Drummond Fielding, Peng Oh, Brent Tan, Viraj Pandya, Ryan Farber, Evan Schneider, and Max Gronke for insightful conversations at the ``Modelling of Multi-phase Gas in Astrophysical Media 2023" conference. DN thanks Chris Rackauckas for useful pointers on using \texttt{Julia}. DN acknowledges funding from NASA 21-ASTRO21-0174.

\nocite{langley00}

\bibliography{example_paper}

\begin{thebibliography}{42}
\providecommand{\natexlab}[1]{#1}
\providecommand{\url}[1]{\texttt{#1}}
\expandafter\ifx\csname urlstyle\endcsname\relax
  \providecommand{\doi}[1]{doi: #1}\else
  \providecommand{\doi}{doi: \begingroup \urlstyle{rm}\Url}\fi

\bibitem[Brunton et~al.(2016)Brunton, Proctor, and Kutz]{Brunton2016}
Brunton, S.~L., Proctor, J.~L., and Kutz, J.~N.
\newblock Discovering governing equations from data by sparse identification of
  nonlinear dynamical systems.
\newblock \emph{Proceedings of the National Academy of Sciences}, 113\penalty0
  (15):\penalty0 3932--3937, 2016.
\newblock \doi{10.1073/pnas.1517384113}.

\bibitem[Chen et~al.(2019)Chen, Rubanova, Bettencourt, and Duvenaud]{Chen2019}
Chen, R. T.~Q., Rubanova, Y., Bettencourt, J., and Duvenaud, D.
\newblock Neural ordinary differential equations, 2019.

\bibitem[{Chevalier} \& {Clegg}(1985){Chevalier} and {Clegg}]{Chevalier1985}
{Chevalier}, R.~A. and {Clegg}, A.~W.
\newblock {Wind from a starburst galaxy nucleus}.
\newblock \emph{\nat}, 317\penalty0 (6032):\penalty0 44--45, Sep 1985.
\newblock \doi{10.1038/317044a0}.

\bibitem[{Cowie} et~al.(1981){Cowie}, {McKee}, and {Ostriker}]{Cowie1981}
{Cowie}, L.~L., {McKee}, C.~F., and {Ostriker}, J.~P.
\newblock {Supernova remnant revolution in an inhomogeneous medium. I -
  Numerical models.}
\newblock \emph{\apj}, 247:\penalty0 908--924, Aug 1981.
\newblock \doi{10.1086/159100}.

\bibitem[{Everett} et~al.(2008){Everett}, {Zweibel}, {Benjamin}, {McCammon},
  {Rocks}, and {Gallagher}]{Everett2008}
{Everett}, J.~E., {Zweibel}, E.~G., {Benjamin}, R.~A., {McCammon}, D., {Rocks},
  L., and {Gallagher}, III, J.~S.
\newblock {The Milky Way's Kiloparsec-Scale Wind: A Hybrid Cosmic-Ray and
  Thermally Driven Outflow}.
\newblock \emph{\apj}, 674:\penalty0 258--270, February 2008.
\newblock \doi{10.1086/524766}.

\bibitem[{Faucher-Giguere} \& {Oh}(2023){Faucher-Giguere} and
  {Oh}]{Faucher-Giguere2023}
{Faucher-Giguere}, C.-A. and {Oh}, S.~P.
\newblock {Key Physical Processes in the Circumgalactic Medium}.
\newblock \emph{arXiv e-prints}, art. arXiv:2301.10253, January 2023.
\newblock \doi{10.48550/arXiv.2301.10253}.

\bibitem[Fronk \& Petzold(2023)Fronk and Petzold]{Fronk2023}
Fronk, C. and Petzold, L.
\newblock Interpretable polynomial neural ordinary differential equations.
\newblock \emph{Chaos: An Interdisciplinary Journal of Nonlinear Science},
  33\penalty0 (4):\penalty0 043101, 2023.
\newblock \doi{10.1063/5.0130803}.

\bibitem[{Gelbrecht} et~al.(2021){Gelbrecht}, {Boers}, and
  {Kurths}]{Gelbrecht2021}
{Gelbrecht}, M., {Boers}, N., and {Kurths}, J.
\newblock {Neural partial differential equations for chaotic systems}.
\newblock \emph{New Journal of Physics}, 23\penalty0 (4):\penalty0 043005,
  April 2021.
\newblock \doi{10.1088/1367-2630/abeb90}.

\bibitem[Goodfellow et~al.(2016)Goodfellow, Bengio, and
  Courville]{Goodfellow2016}
Goodfellow, I., Bengio, Y., and Courville, A.
\newblock \emph{Deep Learning}.
\newblock The MIT Press, Cambridge, MA, 2016.
\newblock ISBN 9780262035613.

\bibitem[{Gronke} \& {Oh}(2018){Gronke} and {Oh}]{Gronke2018}
{Gronke}, M. and {Oh}, S.~P.
\newblock {The growth and entrainment of cold gas in a hot wind}.
\newblock \emph{\mnras}, 480\penalty0 (1):\penalty0 L111--L115, October 2018.
\newblock \doi{10.1093/mnrasl/sly131}.

\bibitem[{Gronke} \& {Oh}(2020){Gronke} and {Oh}]{Gronke2020}
{Gronke}, M. and {Oh}, S.~P.
\newblock {How cold gas continuously entrains mass and momentum from a hot
  wind}.
\newblock \emph{\mnras}, 492\penalty0 (2):\penalty0 1970--1990, February 2020.
\newblock \doi{10.1093/mnras/stz3332}.

\bibitem[Hornik et~al.(1990)Hornik, Stinchcombe, and White]{HORNIK1990}
Hornik, K., Stinchcombe, M., and White, H.
\newblock Universal approximation of an unknown mapping and its derivatives
  using multilayer feedforward networks.
\newblock \emph{Neural Networks}, 3\penalty0 (5):\penalty0 551--560, 1990.
\newblock ISSN 0893-6080.
\newblock \doi{https://doi.org/10.1016/0893-6080(90)90005-6}.
\newblock URL
  \url{https://www.sciencedirect.com/science/article/pii/0893608090900056}.

\bibitem[{Keith} et~al.(2021){Keith}, {Khadse}, and {Field}]{Keith2021}
{Keith}, B., {Khadse}, A., and {Field}, S.~E.
\newblock {Learning orbital dynamics of binary black hole systems from
  gravitational wave measurements}.
\newblock \emph{Physical Review Research}, 3\penalty0 (4):\penalty0 043101,
  November 2021.
\newblock \doi{10.1103/PhysRevResearch.3.043101}.

\bibitem[{Kim} et~al.(2020){Kim}, {Ostriker}, {Fielding}, {Smith}, {Bryan},
  {Somerville}, {Forbes}, {Genel}, and {Hernquist}]{Kim2020}
{Kim}, C.-G., {Ostriker}, E.~C., {Fielding}, D.~B., {Smith}, M.~C., {Bryan},
  G.~L., {Somerville}, R.~S., {Forbes}, J.~C., {Genel}, S., and {Hernquist}, L.
\newblock {A Framework for Multiphase Galactic Wind Launching Using TIGRESS}.
\newblock \emph{\apjl}, 903\penalty0 (2):\penalty0 L34, November 2020.
\newblock \doi{10.3847/2041-8213/abc252}.

\bibitem[{Klein} et~al.(1994){Klein}, {McKee}, and {Colella}]{Klein1994}
{Klein}, R.~I., {McKee}, C.~F., and {Colella}, P.
\newblock {On the Hydrodynamic Interaction of Shock Waves with Interstellar
  Clouds. I. Nonradiative Shocks in Small Clouds}.
\newblock \emph{\apj}, 420:\penalty0 213, January 1994.
\newblock \doi{10.1086/173554}.

\bibitem[{Kopp} \& {Holzer}(1976){Kopp} and {Holzer}]{Kopp1976}
{Kopp}, R.~A. and {Holzer}, T.~E.
\newblock {Dynamics of coronal hole regions. I. Steady polytropic flows with
  multiple critical points.}
\newblock \emph{solphys}, 49\penalty0 (1):\penalty0 43--56, July 1976.
\newblock \doi{10.1007/BF00221484}.

\bibitem[Kuwahara \& Bauch(2023)Kuwahara and Bauch]{Kuwahara2023}
Kuwahara, B.~S. and Bauch, C.~T.
\newblock Predicting covid-19 pandemic waves with biologically and behaviorally
  informed universal differential equations.
\newblock In \emph{medRxiv}, 2023.

\bibitem[{Lamers} \& {Cassinelli}(1999){Lamers} and {Cassinelli}]{Lamers1999}
{Lamers}, H. J.~G.~L.~M. and {Cassinelli}, J.~P.
\newblock \emph{{Introduction to Stellar Winds}}.
\newblock 1999.

\bibitem[LeCun et~al.(2015)LeCun, Bengio, and Hinton]{LeCun2015}
LeCun, Y., Bengio, Y., and Hinton, G.
\newblock \emph{Deep Learning}.
\newblock Nature, New York, NY, 2015.
\newblock \doi{10.1038/nature14539}.
\newblock URL \url{https://www.nature.com/articles/nature14539}.

\bibitem[{Lopez} et~al.(2020){Lopez}, {Mathur}, {Nguyen}, {Thompson}, and
  {Olivier}]{Lopez2020}
{Lopez}, L.~A., {Mathur}, S., {Nguyen}, D.~D., {Thompson}, T.~A., and
  {Olivier}, G.~M.
\newblock {Temperature and Metallicity Gradients in the Hot Gas Outflows of
  M82}.
\newblock \emph{\apj}, 904\penalty0 (2):\penalty0 152, December 2020.
\newblock \doi{10.3847/1538-4357/abc010}.

\bibitem[Lu et~al.(2021)Lu, Mao, Zuo, and Dong]{Lu2021}
Lu, Q., Mao, Z., Zuo, W., and Dong, B.
\newblock Learning nonlinear operators via deeponet based on the universal
  approximation theorem of operators.
\newblock \emph{Nature Machine Intelligence}, 3:\penalty0 270--278, 2021.
\newblock \doi{10.1038/s42256-021-00302-5}.
\newblock URL \url{https://www.nature.com/articles/s42256-021-00302-5}.

\bibitem[Meiss(2007)]{Meiss2007}
Meiss, J.~D.
\newblock \emph{Differential Dynamical Systems}.
\newblock Society for Industrial and Applied Mathematics, 2007.
\newblock \doi{10.1137/1.9780898718232}.
\newblock URL \url{https://epubs.siam.org/doi/abs/10.1137/1.9780898718232}.

\bibitem[{Nguyen} \& {Thompson}(2021){Nguyen} and {Thompson}]{Nguyen2021}
{Nguyen}, D.~D. and {Thompson}, T.~A.
\newblock {Mass-loading and non-spherical divergence in hot galactic winds:
  implications for X-ray observations}.
\newblock \emph{\mnras}, October 2021.
\newblock \doi{10.1093/mnras/stab2910}.

\bibitem[{Nguyen} et~al.(2023){Nguyen}, {Thompson}, {Schneider}, {Lopez}, and
  {Lopez}]{Nguyen2023}
{Nguyen}, D.~D., {Thompson}, T.~A., {Schneider}, E.~E., {Lopez}, S., and
  {Lopez}, L.~A.
\newblock {Dynamics of hot galactic winds launched from spherically-stratified
  starburst cores}.
\newblock \emph{\mnras}, 518\penalty0 (1):\penalty0 L87--L91, January 2023.
\newblock \doi{10.1093/mnrasl/slac141}.

\bibitem[{Pandya} et~al.(2022){Pandya}, {Fielding}, {Bryan}, {Carr},
  {Somerville}, {Stern}, {Faucher-Giguere}, {Hafen}, and
  {Angles-Alcazar}]{Pandya2022}
{Pandya}, V., {Fielding}, D.~B., {Bryan}, G.~L., {Carr}, C., {Somerville},
  R.~S., {Stern}, J., {Faucher-Giguere}, C.-A., {Hafen}, Z., and
  {Angles-Alcazar}, D.
\newblock {A unified model for the co-evolution of galaxies and their
  circumgalactic medium: the relative roles of turbulence and atomic cooling
  physics}.
\newblock \emph{arXiv e-prints}, art. arXiv:2211.09755, November 2022.
\newblock \doi{10.48550/arXiv.2211.09755}.

\bibitem[{Quataert} et~al.(2022{\natexlab{a}}){Quataert}, {Jiang}, and
  {Thompson}]{Quataert2022b}
{Quataert}, E., {Jiang}, Y.-F., and {Thompson}, T.~A.
\newblock {The physics of galactic winds driven by cosmic rays - II. Isothermal
  streaming solutions}.
\newblock \emph{\mnras}, 510\penalty0 (1):\penalty0 920--945, February
  2022{\natexlab{a}}.
\newblock \doi{10.1093/mnras/stab3274}.

\bibitem[{Quataert} et~al.(2022{\natexlab{b}}){Quataert}, {Thompson}, and
  {Jiang}]{Quataert2022a}
{Quataert}, E., {Thompson}, T.~A., and {Jiang}, Y.-F.
\newblock {The physics of galactic winds driven by cosmic rays I: Diffusion}.
\newblock \emph{\mnras}, 510\penalty0 (1):\penalty0 1184--1203, February
  2022{\natexlab{b}}.
\newblock \doi{10.1093/mnras/stab3273}.

\bibitem[Rackauckas \& Nie(2017)Rackauckas and
  Nie]{rackauckas2017differentialequations}
Rackauckas, C. and Nie, Q.
\newblock Differentialequations.jl--a performant and feature-rich ecosystem for
  solving differential equations in julia.
\newblock \emph{Journal of Open Research Software}, 5\penalty0 (1):\penalty0
  15, 2017.

\bibitem[Rackauckas et~al.(2020)Rackauckas, Ma, Martensen, Warner, Zubov,
  Supekar, Skinner, and Ramadhan]{rackauckas2020universal}
Rackauckas, C., Ma, Y., Martensen, J., Warner, C., Zubov, K., Supekar, R.,
  Skinner, D., and Ramadhan, A.
\newblock Universal differential equations for scientific machine learning.
\newblock \emph{arXiv preprint arXiv:2001.04385}, 2020.

\bibitem[{Revels} et~al.(2016){Revels}, {Lubin}, and
  {Papamarkou}]{RevelsLubinPapamarkou2016}
{Revels}, J., {Lubin}, M., and {Papamarkou}, T.
\newblock Forward-mode automatic differentiation in {J}ulia.
\newblock \emph{arXiv:1607.07892 [cs.MS]}, 2016.
\newblock URL \url{https://arxiv.org/abs/1607.07892}.

\bibitem[{Santana} et~al.(2023){Santana}, {Costa}, {Rebello}, {Mafalda
  Ribeiro}, {Rackauckas}, and {Nogueira}]{Santana2023}
{Santana}, V.~V., {Costa}, E., {Rebello}, C.~M., {Mafalda Ribeiro}, A.,
  {Rackauckas}, C., and {Nogueira}, I. B.~R.
\newblock {Efficient hybrid modeling and sorption model discovery for
  non-linear advection-diffusion-sorption systems: A systematic scientific
  machine learning approach}.
\newblock \emph{arXiv e-prints}, art. arXiv:2303.13555, March 2023.
\newblock \doi{10.48550/arXiv.2303.13555}.

\bibitem[{Schneider} \& {Robertson}(2015){Schneider} and
  {Robertson}]{Schneider2015}
{Schneider}, E.~E. and {Robertson}, B.~E.
\newblock {CHOLLA: A New Massively Parallel Hydrodynamics Code for
  Astrophysical Simulation}.
\newblock \emph{\apjs}, 217\penalty0 (2):\penalty0 24, April 2015.
\newblock \doi{10.1088/0067-0049/217/2/24}.

\bibitem[{Smith} et~al.(2023){Smith}, {Fielding}, {Bryan}, {Kim}, {Ostriker},
  {Somerville}, {Stern}, {Su}, {Weinberger}, {Hu}, {Forbes}, {Hernquist},
  {Burkhart}, and {Li}]{Smith2023}
{Smith}, M.~C., {Fielding}, D.~B., {Bryan}, G.~L., {Kim}, C.-G., {Ostriker},
  E.~C., {Somerville}, R.~S., {Stern}, J., {Su}, K.-Y., {Weinberger}, R., {Hu},
  C.-Y., {Forbes}, J.~C., {Hernquist}, L., {Burkhart}, B., and {Li}, Y.
\newblock {Arkenstone I: A Novel Method for Robustly Capturing High Specific
  Energy Outflows In Cosmological Simulations}.
\newblock \emph{arXiv e-prints}, art. arXiv:2301.07116, January 2023.
\newblock \doi{10.48550/arXiv.2301.07116}.

\bibitem[{Stepaniants} et~al.(2023){Stepaniants}, {Hastewell}, {Skinner},
  {Totz}, and {Dunkel}]{Stepanmiants2023}
{Stepaniants}, G., {Hastewell}, A.~D., {Skinner}, D.~J., {Totz}, J.~F., and
  {Dunkel}, J.
\newblock {Discovering dynamics and parameters of nonlinear oscillatory and
  chaotic systems from partial observations}.
\newblock \emph{arXiv e-prints}, art. arXiv:2304.04818, April 2023.
\newblock \doi{10.48550/arXiv.2304.04818}.

\bibitem[{Strickland} \& {Heckman}(2009){Strickland} and
  {Heckman}]{Strickland2009}
{Strickland}, D.~K. and {Heckman}, T.~M.
\newblock {Supernova Feedback Efficiency and Mass Loading in the Starburst and
  Galactic Superwind Exemplar M82}.
\newblock \emph{\apj}, 697:\penalty0 2030--2056, June 2009.
\newblock \doi{10.1088/0004-637X/697/2/2030}.

\bibitem[{Tan} \& {Fielding}(2023){Tan} and {Fielding}]{Tan2023}
{Tan}, B. and {Fielding}, D.~B.
\newblock {Cloud Atlas: Navigating the Multiphase Landscape of Tempestuous
  Galactic Winds}.
\newblock \emph{arXiv e-prints}, art. arXiv:2305.14424, May 2023.
\newblock \doi{10.48550/arXiv.2305.14424}.

\bibitem[Tsitouras(2011)]{TSITOURAS2011}
Tsitouras, C.
\newblock Runge–kutta pairs of order 5(4) satisfying only the first column
  simplifying assumption.
\newblock \emph{Computers \& Mathematics with Applications}, 62\penalty0
  (2):\penalty0 770--775, 2011.
\newblock ISSN 0898-1221.
\newblock \doi{https://doi.org/10.1016/j.camwa.2011.06.002}.
\newblock URL
  \url{https://www.sciencedirect.com/science/article/pii/S0898122111004706}.

\bibitem[Vortmeyer-Kley et~al.(2021)Vortmeyer-Kley, Nieters, and
  Pipa]{Vortmeyer-kley2021}
Vortmeyer-Kley, R., Nieters, P., and Pipa, G.
\newblock A trajectory-based loss function to learn missing terms in
  bifurcating dynamical systems.
\newblock \emph{Scientific Reports}, 11:\penalty0 21181, Oct 2021.
\newblock \doi{10.1038/s41598-021-99609-x}.
\newblock URL \url{https://www.nature.com/articles/s41598-021-99609-x}.

\bibitem[{Wibking} \& {Krumholz}(2022){Wibking} and {Krumholz}]{Wibking2022}
{Wibking}, B.~D. and {Krumholz}, M.~R.
\newblock {QUOKKA: a code for two-moment AMR radiation hydrodynamics on GPUs}.
\newblock \emph{\mnras}, 512\penalty0 (1):\penalty0 1430--1449, May 2022.
\newblock \doi{10.1093/mnras/stac439}.

\bibitem[Wigner(1960)]{Wigner1960}
Wigner, E.~P.
\newblock The unreasonable effectiveness of mathematics in the natural
  sciences.
\newblock \emph{Communications on Pure and Applied Mathematics}, 13\penalty0
  (1):\penalty0 1--14, 1960.

\bibitem[Yin et~al.(2023)Yin, Wu, and Song]{Yin2023}
Yin, S., Wu, J., and Song, P.
\newblock Optimal control by deep learning techniques and its applications on
  epidemic models.
\newblock \emph{Journal of Mathematical Biology}, 87:\penalty0 2121--2148, Apr
  2023.
\newblock \doi{10.1007/s00285-023-01873-0}.
\newblock URL
  \url{https://link.springer.com/article/10.1007/s00285-023-01873-0}.

\bibitem[{Yuan} et~al.(2023){Yuan}, {Krumholz}, and {Martin}]{Yuan2023}
{Yuan}, Y., {Krumholz}, M.~R., and {Martin}, C.~L.
\newblock {The observable properties of cool winds from galaxies, AGN, and star
  clusters - II. 3D models for the multiphase wind of M82}.
\newblock \emph{\mnras}, 518\penalty0 (3):\penalty0 4084--4105, January 2023.
\newblock \doi{10.1093/mnras/stac3241}.

\end{thebibliography}
\bibliographystyle{icml2023}



\end{document}